\documentclass{PoS}

\title{Pion form factor from local-duality QCD sum rule}

\ShortTitle{Pion form factor from local-duality QCD sum rule}

\author{Wolfgang Lucha
\\Institute for High Energy Physics, Austrian
Academy of Sciences, Nikolsdorfergasse 18, A-1050, Vienna,
Austria
\\E-mail:
\email{Wolfgang.Lucha@oeaw.ac.at}}

\author{\speaker{Dmitri Melikhov}
\\Institute for High Energy
Physics, Austrian Academy of Sciences, Nikolsdorfergasse 18,
A-1050, Vienna, Austria, and
\\D.~V.~Skobeltsyn Institute of
Nuclear Physics, Moscow State University, 119991, Moscow,
Russia\\E-mail: 
\email{dmitri\_melikhov@gmx.de}}
\abstract{We present our recent results for the pion elastic form factor \cite{braguta} 
obtained within a local-duality three-point sum rule (a Borel sum rule in the limit of an 
infinite Borel parameter). Our analysis \cite{braguta} includes the $O(1)$ and $O(\alpha_s)$  
contributions and is therefore applicable in a broad range of spacelike momentum transfers. 
Our results demonstrate in essentially model-independent way that the $O(1)$ term, 
which provides the subleading $1/Q^4$ power correction at asymptotically large momentum transfers, 
contributes more than half of the pion form factor in the region $Q^2 \le 20$ GeV$^2$.
To probe the accuracy of local-duality sum rules for form factors, we apply precisely the same procedures 
to extract the form factor in a quantum-mechanical potential model. 
Comparison of the exact form factor known in this model 
with the result of the sum-rule calculation gives a probe of the systematic error of the method.    
In our example this error is found to be at the level of 10--20\%. We expect similar  
systematic errors for form factors obtained from local-duality sum rules in QCD.}

\FullConference{8th Conference Quark Confinement and the Hadron Spectrum\\
         September 1--6, 2008\\
         Mainz, Germany}

\begin{document}
\section{Introduction}
The values of hadron parameters (in particular, form factors) extracted from QCD sum rules 
depend on two ingredients: (i) the field-theoretic calculation of the relevant correlator  
and (ii) the technical ``extraction procedure'', which is external to the underlying field theory. 
The second ingredient introduces a systematic error which is very hard to control in any 
version of QCD sum rules \cite{lms_sr}. 
So, in order to probe the accuracy of the sum-rule 
predictions it is plausible to apply different versions of sum rules to the same quantities 
and compare the results. 

In this talk we report our recent analysis of the pion form factor \cite{braguta} 
within the so-called local-duality (LD) version of three-point QCD sum rules 
(for details of this method see \cite{radyushkin} and references therein) and try to 
get an idea of the expected accuracy of this calculation. 

The LD sum rules are the Borel sum rules in the limit of an infinitely large Borel parameter.   
For the relevant choice of the pion interpolating current, the condensate contributions to the correlators 
vanish in this limit and the pion observables are given by dispersion integrals via the spectral densities 
of purely perturbative QCD diagrams. The integration region in the dispersion integrals is restricted 
to the pion ``duality interval''. 

This approach has the following attractive features: 
(i) it is applicable in a broad range of momentum transfers, and 
(ii) it does not refer directly to the pion distribution amplitude. 
Therefore, it allows us to study in a relatively model-independent way the interplay 
between perturbative and nonperturbative dynamics in the pion from factor at intermediate momentum transfers. 
\section{Sum rule}
Let us consider the sum rules for the pion form factor and the decay constant in the LD limit, where 
all condensate contributions vanish. 
According to the standard assumption that the ground-state 
contribution is dual to the low-energy region of the free-quark diagrams, for the case of massless quarks
we obtain to $\alpha_s$ accuracy  
\begin{eqnarray}
\label{ldff}
F^{\rm LD}_{\pi} (Q^2)(f^{\rm LD}_\pi)^2 &=& 
\frac{1}{\pi^2}\int_0^{s_0}ds_1\int_0^{s_0}ds_2 
\left[\Delta^{(0)} (s_1, s_2, Q^2)+ \alpha_s \Delta^{(1)} (s_1, s_2, Q^2)+O(\alpha_s^2)\right], 
\\ 
\label{ldfpi}
(f^{\rm LD}_\pi)^2 &=& 
\frac{1}{\pi}\int_0^{s_0}ds\left[\rho^{(0)}(s)+\alpha_s\rho^{(1)}(s)+O(\alpha_s^2)\right]. 
\end{eqnarray}
The double dispersion representation (\ref{ldff}) for $F_\pi f_\pi^2$, even in the LD limit, has two 
essential ambiguities: (a) The choice of the shape of the duality region in the $s_1$--$s_2$ plane: 
the simplest choice, implemented in (\ref{ldff}), is a square, but any other region symmetric under 
$s_1\leftrightarrow s_2$ may be chosen. 
(b) The duality interval $s_0$ in the 3-point correlator may (and should) depend on $Q^2$. 
This dependence has been neglected. 
Now, we remind the reader that sum rules are predictive {\it only} if one imposes a criterion 
to fix the effective continuum threshold \cite{lms_sr}.
In (\ref{ldff}) we have made an additional essential assumption --- we have set  
the parameter $s_0$ to be the same for both sum rules (\ref{ldff}) and (\ref{ldfpi}). 
The form factor (\ref{ldff}) has the following attractive properties: 

\noindent (i) 
It satisfies the normalization condition $F_\pi(Q^2=0)=1$ due to the vector Ward identity, 
which provides the relation between the spectral density of the self-energy diagram and the double 
spectral density of the triangle diagram at zero momentum transfer:  
\begin{eqnarray}
\label{wi}
\lim_{Q^2\to 0}\Delta^{(i)}(s_1,s_2,Q^2)=\pi \rho^{(i)}(s_1)\delta(s_1-s_2),
\qquad \rho^{(0)}(s)=\frac{1}{4\pi},
\qquad \rho^{(1)}(s)=\frac{1}{4\pi^2}. 
\end{eqnarray}
Clearly, for consistency one should then take into account the radiative corrections 
to the same order in the sum rules for two- and three-point correlators. 

\noindent (ii) 
For $Q^2\gg s_1, s_2$, one finds $\Delta^{(0)}(s_1,s_2,Q^2)\to
{3(s_1+s_2)}/{2 Q^4}$, $\Delta^{(1)}(s_1,s_2,Q^2)\to {1}/{2\pi Q^2}$ \cite{braguta}. 
Substituting these relations into (\ref{ldff}) and identifying $f^{\rm LD}_\pi$ with the pion decay 
constant $f_\pi$ yields at large $Q^2$:
\begin{eqnarray}
\label{norunning}
F^{\rm LD}_\pi(Q^2)=\frac{8\pi f_\pi^2\alpha_s}{Q^2}+\frac{96\pi^4 f_\pi^4}{Q^4}
+O\left(\alpha_sf_\pi^4/Q^4\right)+O\left(\alpha_s^2\right).
\end{eqnarray}
Interestingly, we have obtained the correct asymptotic behaviour of the pion 
form factor in the limit $Q^2\to\infty$ 
(up to the running of $\alpha_s$). 
This property is due to (i) the factorization of the $O(\alpha_s)$ cut triangle diagram at large $Q^2$ 
and (ii) choosing the same value of $s_0$ in 2- and 3-point correlators. 

However, there are also obvious problems:   

First, the sum rule cannot be directly applied at small $Q^2$, although the expression (\ref{ldff}) 
leads to the correct normalization of the form factor: 
Recall that the OPE for the three-point correlator was obtained in the region where 
all three external variables $|p_1^2|$, $|p_2^2|$, and $Q^2$ are large. 
[A technical indication of the fact that the LD sum rule (\ref{ldff}) cannot be applied at very small $Q^2$ 
is the presence of terms $\sim \sqrt{Q^2}$ leading to an infinite value of the pion radius.] 
Moreover, the spectral density $\Delta(s_1,s_2,Q^2)$ contains the terms $O(1)$ and 
$O(\alpha_s)$, whereas higher powers are unknown. Since the 
coupling constant $\alpha_s$ is not small in the soft region,
our spectral density is not sufficient for application to the form factor 
at $Q^2\le 1$ GeV$^2$. 

Second, in order to apply the obtained formulas for large $Q^2$, higher-order radiative corrections, 
leading to the running of $\alpha_s$, should be taken into account. Such an accuracy is beyond our 
two-loop calculation; nevertheless, a self-consistent expression for 
the form factor applicable for all $Q^2>0$ may be obtained from (\ref{ldff}) by using in the sum rule for 
$F_\pi$ the $Q^2$-dependent threshold 
\begin{eqnarray}
\label{threshold3}
s_0(Q^2)=\frac{4\pi^2f_\pi^2}{1+{\alpha_s(Q^2)}/{\pi}}. 
\end{eqnarray}
One should understand the meaning of $\alpha_s(Q^2)$ at small $Q^2$: in \cite{radyushkin} 
it was argued that $s_0$ is the relevant scale of $\alpha_s$ in the LD sum rules for the decay 
constant and for the form factor at $Q^2=0$. 
For the calculations we have assumed the freezing of  $\alpha_s$ at the value 0.3.  

\begin{figure}[t]
\begin{center}
\begin{tabular}{cc}
\includegraphics[width=7cm]{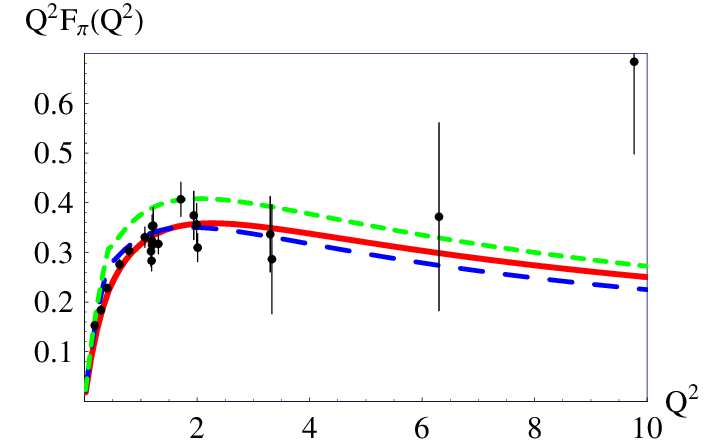}&\includegraphics[width=7cm]{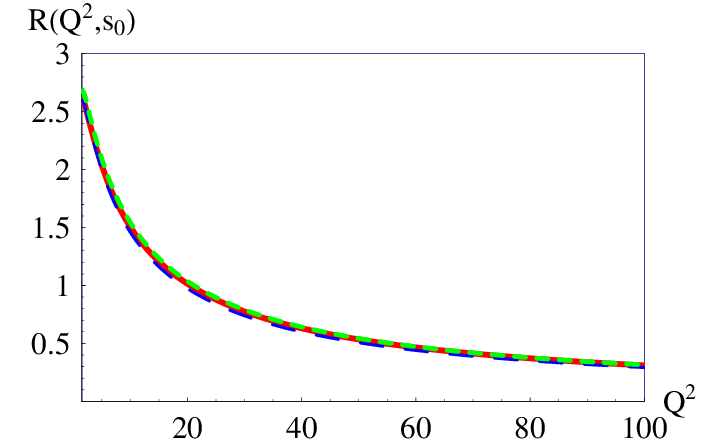}\\
(a)& (b)
\end{tabular}
\caption{\label{Plot:1}
(a) The pion form factor vs $Q^2$ (in units of $GeV^2$) for $Q^2\ge 0.5$ GeV$^2$. 
Experimental data from \cite{data_largeQ2}. 
(b) The ratio of the $O(1)/O(\alpha_s)$ contributions to the form factor. 
Solid (red) line: the result of the calculation according to (\protect\ref{ldff});   
short-dashed (green) line: the form factor obtained with constant $s_0=0.65$ GeV$^2$; 
long-dashed (blue) line: $s_0=0.6$ GeV$^2$.}
\end{center}
\end{figure}

The results for the pion form factor are shown in Fig.~\ref{Plot:1}a, and 
Fig.~\ref{Plot:1}b presents the ratio of the $O(1)$ and the $O(\alpha_s)$ contributions to the form
factor vs $Q^2$ for different models of the effective continuum thresholds. One can clearly see 
that the ratio is mainly determined by the corresponding double spectral densities 
$\Delta^{(0)}$ and $\Delta^{(1)}$, whereas its sensitivity to the effective continuum threshold is rather weak. 

Obviously, the $O(1)$ term, which provides the subleading $1/Q^4$ power correction 
at large $Q^2$, dominates the form factor at low $Q^2$, and still gives 50\% at $Q^2=20$ GeV$^2$. The
$O(\alpha_s)$ term gives more than 80\% of the form factor only above $Q^2=100$ GeV$^2$. 
Such a pattern of the pion form factor behaviour has been conjectured many times in the literature; 
we now obtain this behaviour in an explicit calculation. This analysis supports the results 
of \cite{anis} obtained by inclusion of $O(1)$ and $O(\alpha_s)$ terms within the dispersion approach. 

\begin{figure}[hb]
\begin{center}
\begin{tabular}{cc}
\includegraphics[width=8cm]{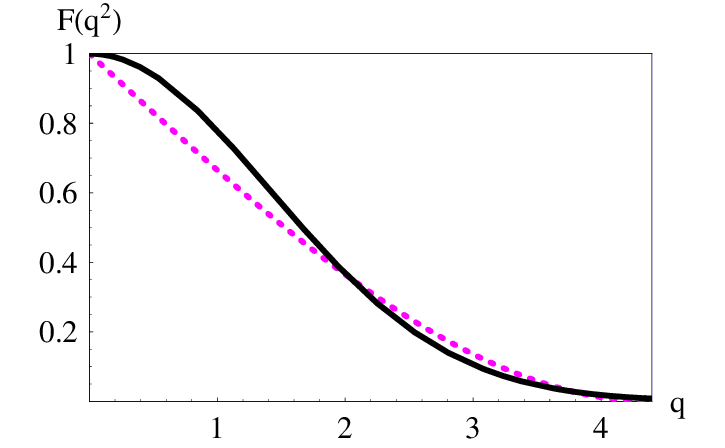}
\end{tabular}
\caption{\label{Plot:2}
The form factor obtained from the LD sum rule in a harmonic-oscillator quantum-mechanical model.  
Solid line: exact form factor, dotted line: the form factor obtained from the LD sum rule with a 
constant effective continuum threshold fixed from the known value of the decay constant 
of the ground state.}
\end{center}
\end{figure}
\section{Discussion and Conclusions}
We presented the results of the first analysis of the pion form factor which takes into account both the 
$O(1)$ and the $O(\alpha_s)$ contributions within the LD sum rule \cite{braguta}. These ingredients are 
crucial for the possibility to consider the form factor in a broad range of $Q^2$ and to study   
the transition from the nonperturbative to the perturbative region. 

Let us summarize the essential ingredients and the lessons to be learnt from our analysis:  

\noindent
1. We have included the exact $O(1)$ and $O(\alpha_s)$ terms 
into the spectral representation for the form factor, and omitted the $O(\alpha_s^2)$ terms, 
which are expected to contribute at the level of less than $\sim 10$\%. 

\noindent
2. The numerical results for the form factor from the LD sum rule depend 
sizeably on the model for the effective continuum threshold 
used for the calculations: this very quantity determines to a great extent the value 
of the extracted form factor. 
The possibility to fix this effective continuum threshold is the weak point of the 
approaches based on all versions of 
sum rules, which limits their predictivity \cite{lms_sr}. 
We used the same threshold in the two- and three-point sum rules considered;    
this allows us to relate the value of the threshold to the pion decay constant, known experimentally. 
In this way we have no free {\it numerical} parameters in our analysis. However,  
our analysis is obviously not free from systematic uncertainties.   

\noindent
3. One cannot assign any rigorous error to the obtained form factor. 
In order to get an idea of the corresponding accuracy, we have used precisely the same algorithms 
within the non-relativistic harmonic-oscillator model, where the exact form factor may be calculated 
independently, and found the error to be at a 10--20\% level (see Fig.~\ref{Plot:2}). 
We believe a similar estimate to be true for our QCD calculation. 

\noindent
4. We can control rather well the relative weights of the $O(1)$ and $O(\alpha_s)$ contributions to 
the form factor: their ratio is practically independent of the continuum threshold.  
Thus, our results convincingly show that the $O(\alpha_s)$ contribution to the pion form 
factor remains below 50\% at $Q^2\le 20$ GeV$^2$; 
these results definitely speak against the pQCD approach to exclusive processes, which assumes  
the form factor to be described by the $O(\alpha_s)$ term already at $Q^2$ of the order of several GeV$^2$. 

\noindent 
\acknowledgments
We are grateful to Victor Braguta for collaboration on this subject and to Alexander Bakulev 
and Silvano Simula for valuable discussions. D.~M.~was supported by the Austrian Science Fund 
(FWF) under project P20573 and by RFBR under project 07-02-00551. 

\vspace{-.2cm}

\end{document}